\documentstyle[prb,aps]{revtex}

\begin {document}


\title {Growth model with restricted surface relaxation}

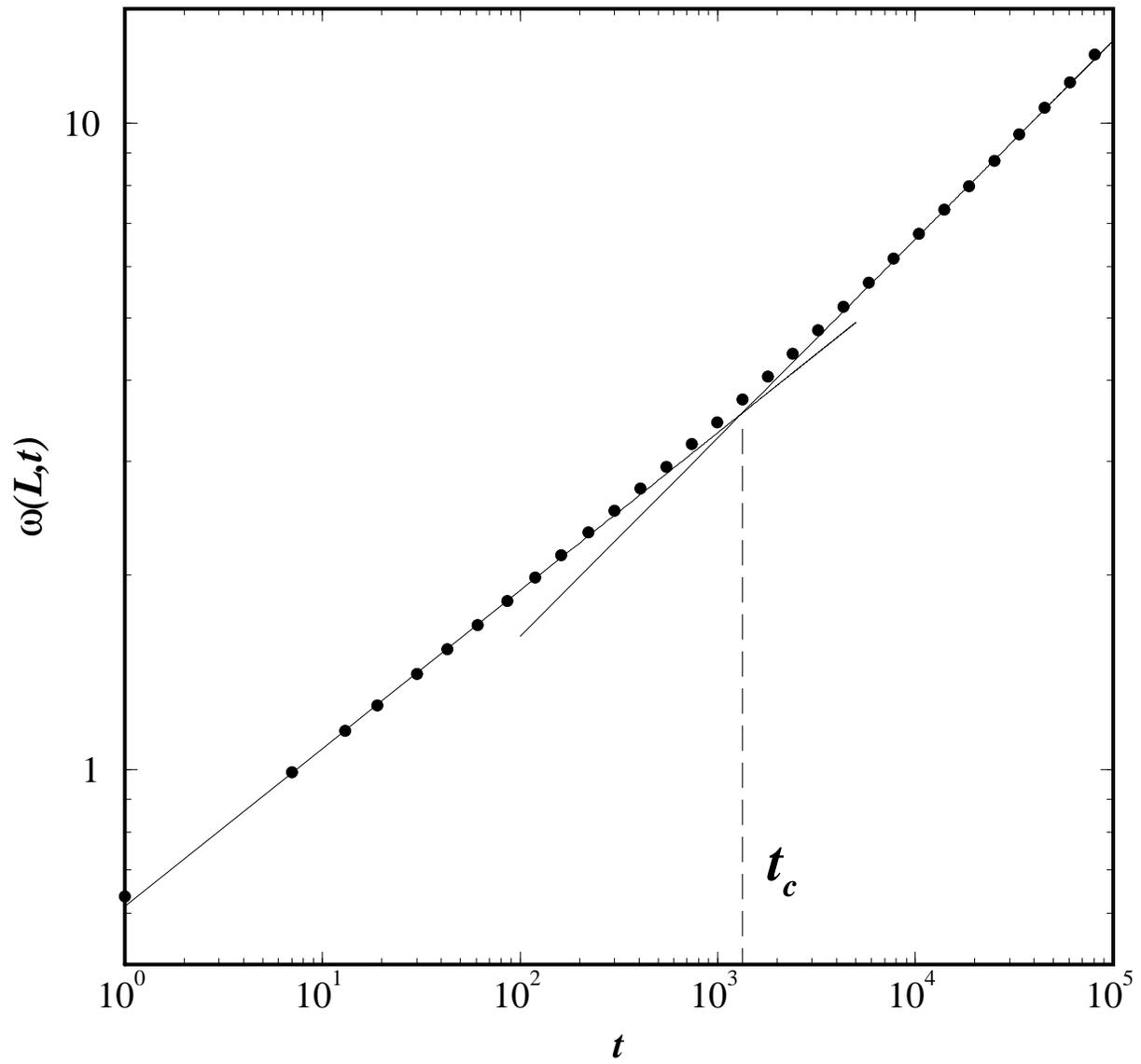
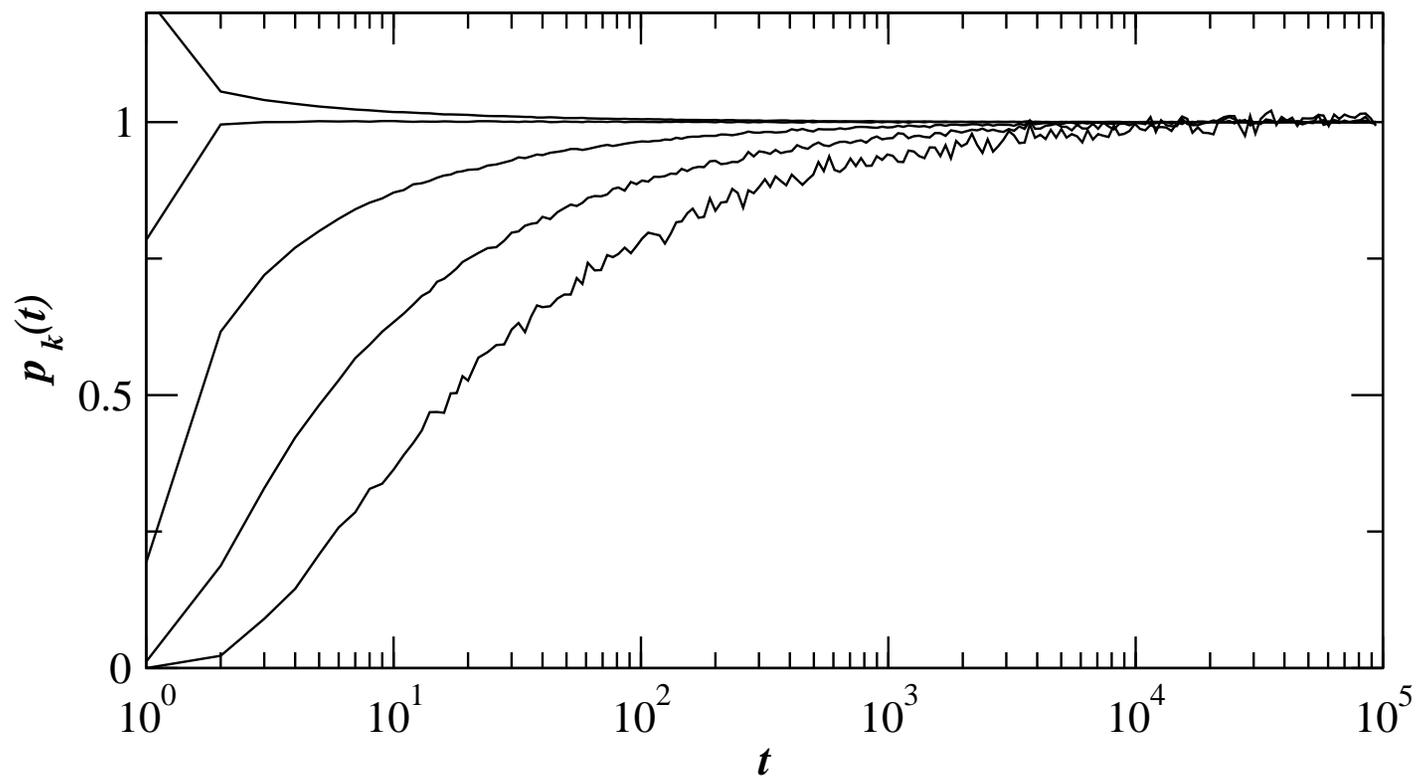
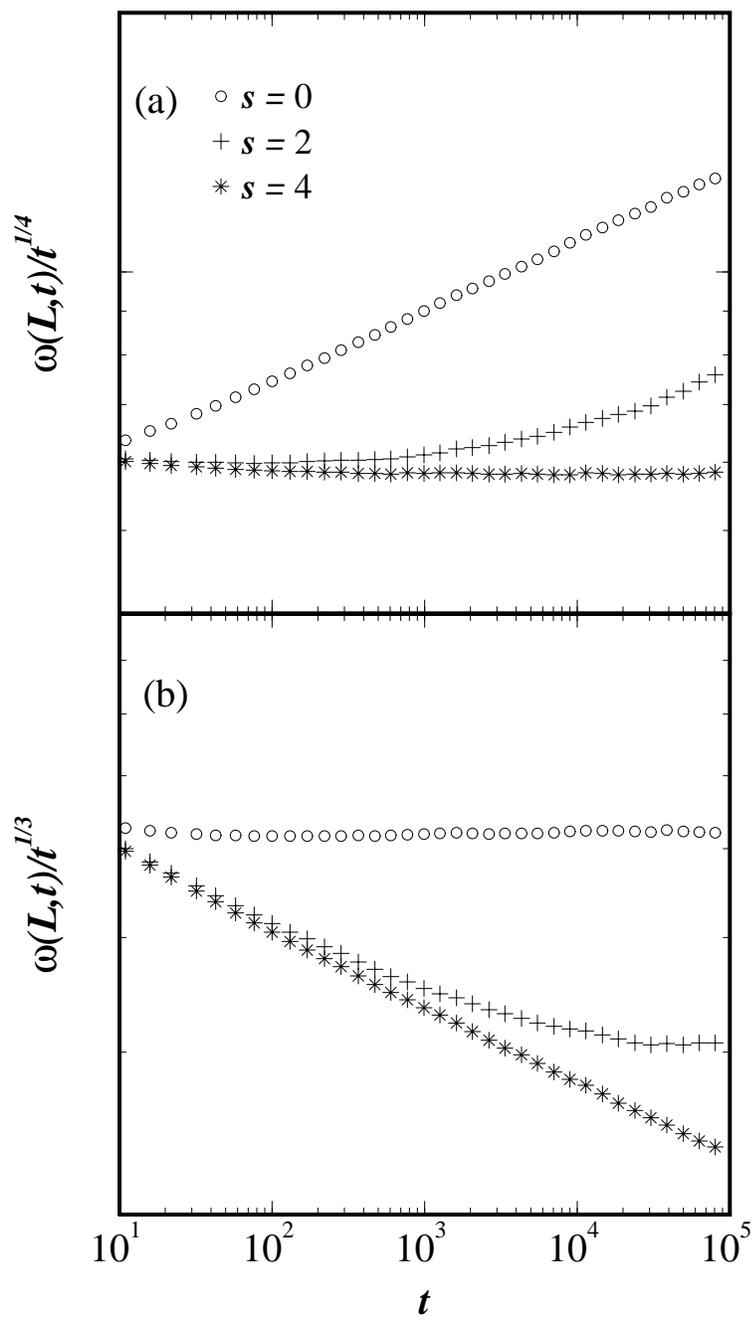
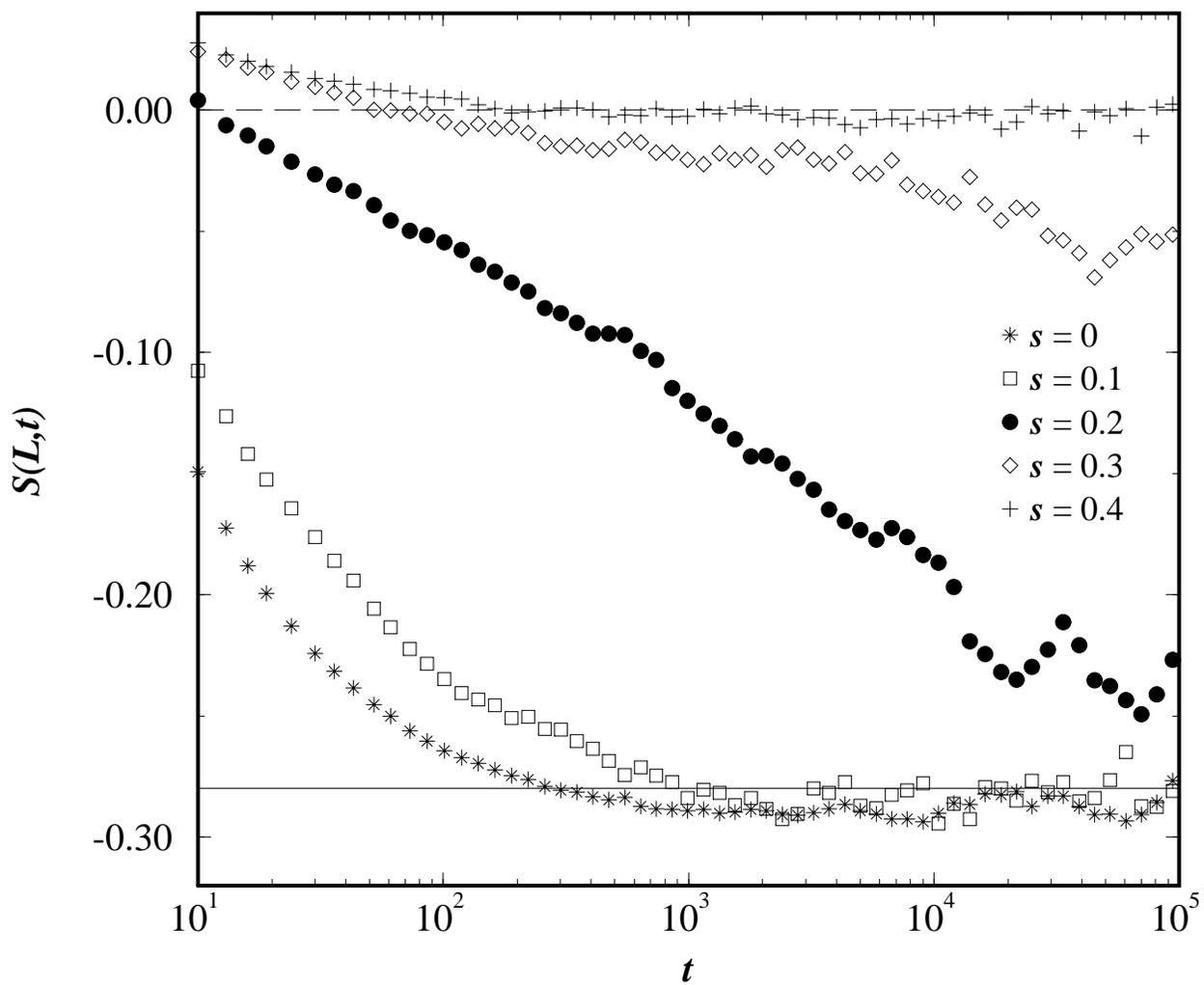
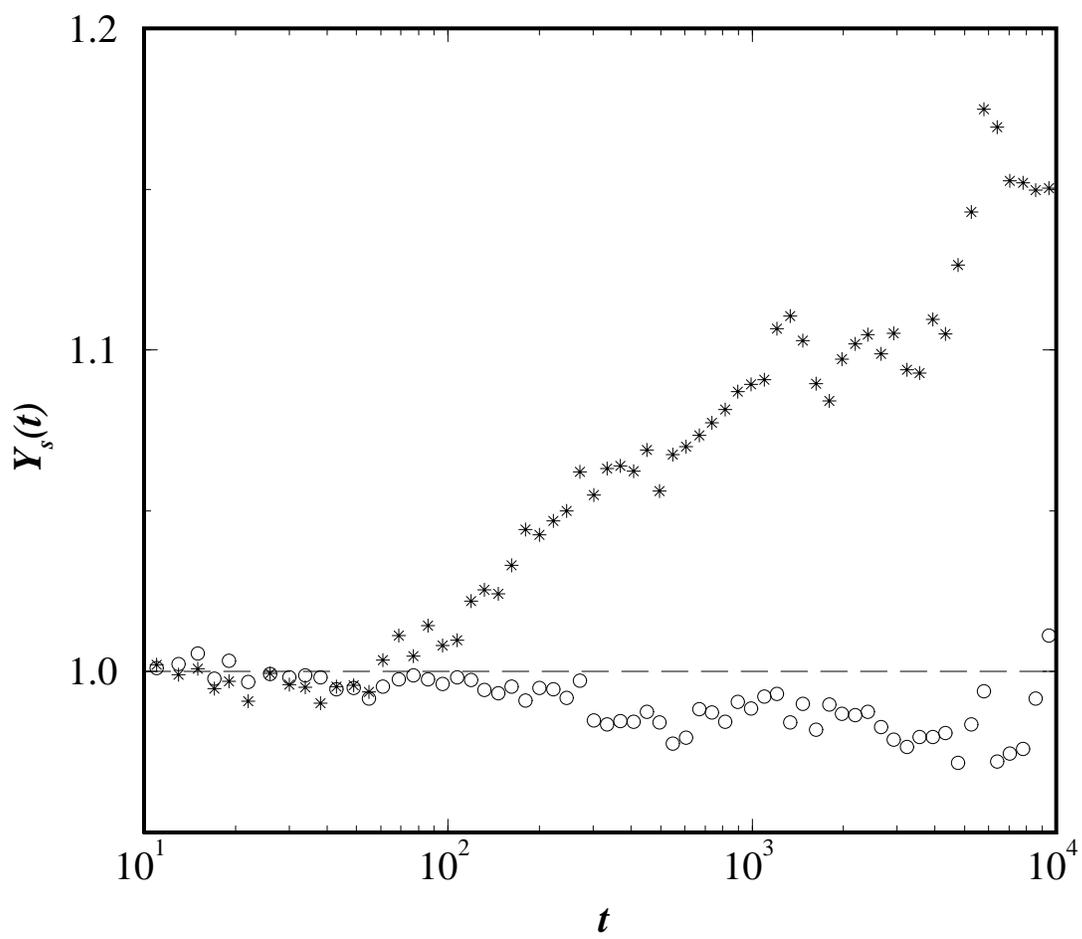
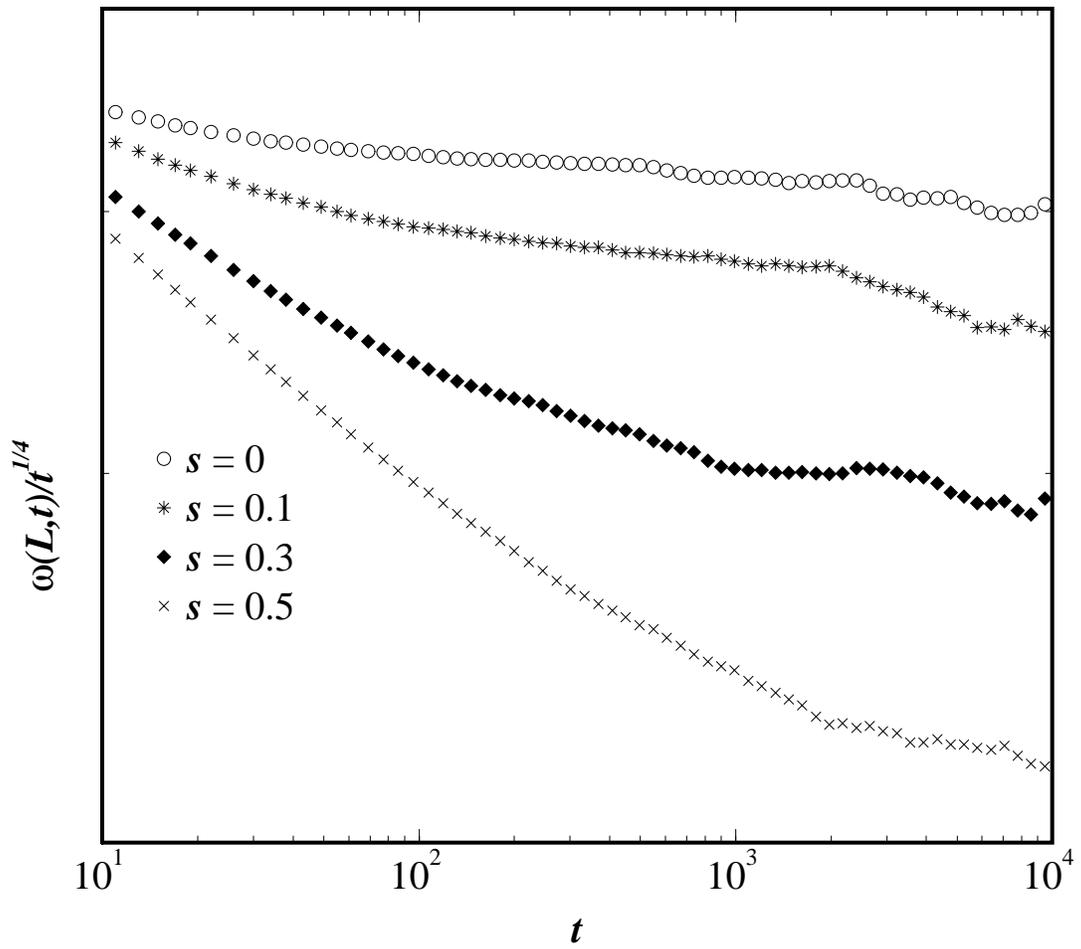
\author {T. J. da Silva and J. G. Moreira} 
\address {Departamento de F\'{\i}sica, Instituto de Ci\^encias Exatas,\\
Universidade Federal de Minas Gerais, Caixa Postal 702,\\ 
30161-970, Belo Horizonte, MG - Brazil} 

\date {\today}
\maketitle

\begin {abstract}
We simulate a growth model with restricted surface relaxation process
in $d=1$ and $d=2$, where $d$ is 
the dimensionality of a flat substrate. In this model, each particle
can relax on the surface to a local minimum, as the Edwards-Wilkinson 
linear model, but only within a distance $s$. 
If the local minimum is out from this distance, the particle
evaporates through a refuse mechanism similar to the Kim-Kosterlitz
nonlinear model. In $d=1$,
the growth exponent $\beta$, measured from the temporal behavior
of roughness, indicates that in the coarse-grained limit, the linear term
of the Kardar-Parisi-Zhang equation dominates in short times 
(low-roughness) and, in asymptotic times, the nonlinear term prevails.
The crossover between linear and nonlinear behaviors occurs in a
characteristic time $t_c$ which only depends on the magnitude of the
parameter $s$, related to the nonlinear term.
In $d=2$, we find indications of a 
similar crossover, that is, logarithmic temporal behavior of roughness in short
times and power law behavior in asymptotic times.

\end {abstract}
\narrowtext 
\twocolumn[\hsize\textwidth\columnwidth\hsize\csname @twocolumnfalse\endcsname
\vskip2pc]
\newpage
%
%
%
%
%
%
\section {INTRODUCTION}
Self-affine interfaces generated by nonequilibrium surface growth
have been intensively studied in recent years [1-3]. Kinetic roughening models 
as
ballistic deposition \cite{db}, Eden model \cite{eden}, solid-on-solid model 
(SOS) with surface
relaxation \cite {ew,family86}, SOS with refuse \cite{kk} and SOS with 
diffusion 
\cite {wolfvillain} are some
examples of growth models that belongs to distinct universality classes. 
Theoretically, these
growth processes are studied in three different schemes: through a continuum
description using Langevin-like equation and renormalization group analisys
for solving it; through numerical solutions of these equations; and
through computer simulations of discrete models.
The main goal is to obtain the universality class of a specific model and to 
get 
information about the presence of nonlinearities and broken symmetries.

In computer simulation of lattice growth models, interfaces are described by
a discrete set $\{h_i(t)\}$ which represents the height of a site 
$i$ at the time $t$. Such an interface has $L^d$ sites, where 
$L$ is the linear size and $d$ is the dimension of the substrate.
The
roughness of the interface $\omega$ is defined as the root mean square of the
$\{h_i-{\bar h}\}$ distribution,
	\begin{equation}
	\omega^2(L,t)=\left<{1\over L^d}\sum_{i=1}^{L^d}(h_i-\bar h)^2\right>,
	\end{equation}
where $\bar h$ is the mean height at time $t$ and the angular
brackets means the average over independent samples.

The universality class of a discrete growth model is obtained through
the temporal and spatial behaviors of
roughness. In the most of kinetic roughening processes, which
starts at $t=0$ from a flat substrate, the temporal
behavior of roughness is described by the power law behavior,
$\omega(L,t)\sim t^{\beta}$, when $1\ll t\ll t_{\times}$, and its spatial behavior, 
in the steady state, is described by 
$\omega_{sat}(L)\sim L^{\alpha}$, for
$t\gg t_{\times}$. The exponents $\beta$ and $\alpha$ are the growth and roughness
exponents, respectivelly, and $t_{\times}$ is the saturation time.
These two behaviors are joined at the Family and Vicsek dynamical scaling
relation\cite{fv} 
	\begin{equation}
	\omega(L,t)\sim L^{\alpha}f\left({t\over L^z}\right),
	\end{equation}
where the function $f(x)$ must be $L$-independent. This function
scales as $f(x)\sim x^{\beta}$ for short times and tends to a constant in
the steady state. The dynamical exponent $z$ is related with $\alpha$ and
$\beta$ through the relationship $z=\alpha/\beta$ and it shows how the
saturation time depends on the system size $L$: $ t_{\times}\sim L^z$. Two of
these exponents predict which universality class a model belongs to.

These universality classes are related to the dominant term of the stochastic
differential equation in the continuum limit. In the next section, we describe two
stochastic differential equations which represent two distinct universality classes.
The first is a linear equation proposed, in 1982, by Edwards and Wilkinson
(EW equation) \cite {ew}. The second is a nonlinear equation, introduced by 
Kardar, Parisi and Zhang in 1986 (KPZ equation) \cite {kpz}. When the
nonlinear term of this equation is null, the EW linear equation is recuperated.

In this article, we report on simulations of a growth model with
restricted relaxation process (called RR model), which was proposed
in order to study the crossover between the linear and nonlinear regimes 
of the KPZ equation. The RR combines
features of the Edwards-Wilkinson (EW) model \cite {ew,family86}, related to
the EW linear equation, and the  
Kim-Kosterlitz (KK) model \cite{kk}, related to the KPZ nonlinear equation. 
The crossover between linear and nonlinear regimes of the
KPZ equation was studied numerically in $d=1$ and $d=2$ 
through numerical solutions of this equation where variations of the amount
of nonlinearity was allowed. Moser and Kert\'esz \cite{moser} 
found, in $d=1$, the growth exponent $\beta=1/3$ for all values of the 
nonlinearity. In $d=2$, the authors have found $\beta=0.240$, close to 
the Kim and 
Kosterlitz value \cite {kk}. So, a crossover between linear 
and nonlinear regimes was not verified. Grossmann, Guo and Grant \cite {ggg} 
also have obtained numerical solutions of KPZ equation where the surface 
tension was fixed and the nonlinear parameter was continuously changed. 
For small systems, the authors have found growth exponents in the interval
$1/4< \beta <1/3$. So, the crossover was characterized by
a continuum change in the growth exponent, which in $d=1$ means that the
dynamical exponent $z$ depends on the amount of the nonlinearity. 
This continuous crossover in $d=1$ was also 
verified by simulations on growth models \cite {yks,pelle}. Theoretically, 
in $d=1$, 
Nattermann and Tang \cite {nattang} have studied the KPZ equation in the
low nonlinear limit through renormalization group analisys and they have
obtained a different result: two behaviors separated by a characteristic 
time $t_c$. 
For $t\ll t_c$, the linear behavior of roughness was found, while for $t\gg t_c$,
the nonlinearity dominates.

This article is organized in sections. The next section presents the
general approach for kinetic roughening and describes the EW and
KK models. In section III, we introduce the RR model and we show our
results which confirm the theoretical previsions of
Natterman and Tang \cite {nattang} and, in section IV, we
finally show our conclusions.
\section {THEORY AND DISCRETE MODELS}
In the continuum (coarse-grained) description, the interface motion is
described through Langevin-like equations [1-3],
	\begin{equation}
	{\partial h({\bf x},t)\over\partial t}=v_0+\eta({\bf x},t)+
	\Phi\left[{h({\bf x},t)}\right]~~.
	\end{equation}
In this equation, $v_0$ and $\eta({\bf x},t)$ are the deposition rate 
and its noise, respectively.
This white noise has zero mean and variance given by
	\begin{equation}
	\left<\eta({\bf x},t)\eta({\bf x}^{'},t^{'})\right>=
	 D\delta^{d}({\bf x} -{\bf x}^{'})\delta(t-t^{'}).
	\end{equation} 
$\Phi\left[{h({\bf x},t)}\right]$ is the term related to the 
correlations between neighbors which can have linear and nonlinear functions
of $h({\bf x},t)$.

Our interest is focused on the SOS model with
surface relaxation \cite {ew,family86} and at the SOS model with 
restriction \cite {kk}. 
The SOS model with surface relaxation, introduced by Edwards and
Wilkinson \cite{ew} in 1982, is a random deposition of particles where 
the difference of height constraint between the neighbors $\{j\}$ of a site
$i$ is given by 
	\begin{equation}
	h_i-h_{\{j\}}< M,
	\end{equation}
where $M$ is the parameter that controls the roughness. In this work, we
always use $M=1$.
If the height of the deposited particle on the site $i$
does not satisfy the height constraint, this particle must be
moved to a local miminum. Family \cite{family86} have obtained for this model, 
in $d=1$, the exponents $\beta=0.25(1)$ and $\alpha=0.48(2)$. This result indicates
that this model, in a coarse-grained limit, belongs to the universality
class defined by the EW equation
	\begin{equation}
	{\partial h({\bf x},t)\over\partial t}=v_0+\eta({\bf x},t)+
	\nu\nabla^{2}h({\bf x},t),
	\end{equation}
where 
the laplacian term is related to the surface relaxation process.
The exponents of this linear equation, obtained through Fourier 
analysis \cite{ew,nattang},
are $\beta=1/4$, $\alpha=1/2$ and $z=2$ for $d=1$.
For $d=2$, which is the critical dimension of Eq.6,
these exponents are $\beta=\alpha=0$ which means that the roughness has 
logarithimic behavior in space and time ($\omega^2\sim\log t$, for $t\ll L^z$,
and $\omega^2_{sat}\sim\log L$, for $t\gg L^z$). Both, the EW model and the
EW equation, generate gaussian height distributions as well.

In the SOS model with restriction, particles are also randomly deposited
onto a substrate and the difference of height constraint is the same of
the EW model, Eq.5, but any kind of relaxation is allowed.
If the height of a deposited particle does not satisfy Eq.5, this choice
must be refused, that is, the particle evaporates.
This model was proposed in 1989 by Kim and Kosterlitz \cite{kk}
in order to study nonlinear kinetic roughening in high dimensions.
They numerically showed, in $d=1$, that this model, named KK model,
belongs to the universality class of
the well known KPZ equation, proposed by Kardar, Parisi and Zhang \cite {kpz} 
	\begin{equation}
	{\partial h({\bf x},t)\over\partial t}=v_0+\eta({\bf x},t)+
          \nu\nabla^{2}h({\bf x},t)+{
          \lambda\over 2}\left[\nabla h({\bf x},t)\right]^{2}.
	\end{equation}
The appearance of the nonlinear term 
$\left[\nabla h({\bf x},t)\right]^{2}$ is due to the lateral
growth, that is, the dependence of the growth velocity on a local
normal of the growing interface,
or to the appearance of a perpendicular driven force that leads to a growth 
velocity 
greater or smaller than the deposition rate $v_0$. In the case of the
KK model, for example, the refuse mechanism makes the growth velocity
smaller than the deposition rate.
In $d=1$, the exponents of this equation \cite{kpz} are 
$\beta=1/3$, $\alpha=1/2$ and $z=3/2$. In $d=2$, the analytical solution is 
not known.
In $d=1$, numerical simulations of the KK model \cite {kk} indicate
$\beta=0.332(5)$ and the $\alpha$-exponent close to the expected value 
$(\alpha=1/2)$ and, in $d=2$,
$\beta=0.250(5)$ and $\alpha=0.40(1)$.

Eq.7 is not invariant under the $h\rightarrow -h$ transformation, which
means that the up-down symmetry is broken in surfaces generated by a
KPZ process. This fact leads to deviations in the gaussian caracter of
the height distributions which can be measured using other moments of
the distribution. 
The Eq.1 can be generalized for any moments of height distribution as
	\begin{equation}
	W_q(L,t)=
	\left<{1\over L^d}\sum_{i=1}^{L^d}(h_i-\bar h)^q\right>~~,
	\end{equation}
where $q$ is the order of the moment. Note that the roughness
$\omega(L,t)$ is related to the second moment: $\omega^2(L,t)=W_2(L,t)$.	
A growing profile has up-down symmetry when positive and negative local
curvatures are equals, and, in this case, $W_3(L,t)$ vanishes. On the other
hand, when asymmetries are present, $W_{3}(L,t)\neq 0$. The skewness,
defined by
\begin{equation}
	S(L,t)={W_3(L,t)\over W_2^{3/2}(L,t)},
\end{equation}
is the ideal measurement of deviations from gaussian behavior. In the 
case of the EW model, $S(L,t)=0$ always. 
For systems in KPZ class in $d=1$, Krug {\it et al.} \cite {krugskew} have indicated 
$|S(L,t)|\approx 0.28$ as an universal value in the transient state. In the
steady state, the profile shows a random-walk character, that is, the height
distribution is gaussian and $S_{sat}(L,t\gg t_{\times})=0$.
In $d=2$,
numerical simulations of the KK model indicate $S\approx -0.40$, in the
transient state, and $S_{sat}\approx 0.28$, in the steady state \cite {landau2,skew}.
\section {MODEL DESCRIPTION AND RESULTS}
In this article, we report on simulations of the growth model with
restricted surface relaxation model (RR model). In the EW model, each 
incoming particle must
search the local minima when the height constraint (Eq.5) 
is not satisfied. We introduce a parameter
$s$ that is the number of lattice units allowed for the relaxation process.
If the deposited particle does not find the minimum after $s$ relaxations,
then this choice must be refused, as in KK model. For $s=0$, the KK model is
recuperated and $s\rightarrow\infty$ yields the EW model.
\subsection {$d=1$ Results}
Figure 1 shows the log-log plot of roughness $\omega(L,t)$ vs time $t$ 
for a system with $L=10^5$ sites and $s=2$. The time unity means $L^d$ 
attempts of deposition.
The two straight lines in the figure are showing 
the power-law fits with $\beta=0.249(1)$ $(1<t<10^3)$ and $\beta=0.332(1)$ 
$(10^3<t<10^5)$. The intersection of these two lines define the crossover time 
$t_c$. This crossover is easy to understand considering the increase of
roughness: in short times, when the roughness is still small, particles do
not need to relax very much for seaching local minima, so the linear
behavior might dominate.
In large times, on the other hand, we note the appearance of relaxation
lenghts bigger than those observed in short times. As relaxation processes are
linked to refuse processes in this model, the system undergoes a crossover 
to the nonlinear behavior. The crossover time $t_c$ is independent of the 
system size $L$ and it is only a function of the parameter $s$.

For a better understanding of this crossover we study the
statistics of relaxations in the EW model, where relaxations of all
sizes can occur.
At the time $t$, be
$\left<N_{k}(t)\right>$ the mean number of particles which diffused $k$
sites searching local minima. So, $k$ relaxations occur with
probability
\begin {equation}
P_k(t)={\left<N_{k}(t)\right>\over L^d}~~.
\end {equation}
These probabilities have an initial temporal dependence and a
steady state $P_k(\infty)$, whose values are shown in Table 1. 
We note a strong decrease of the probability with the number 
of sites diffused, which shows the rare ocurrences of
the relaxation with large relaxation lenghts.

It is interesting to analyse the approach of
this probability to the steady state because each $P_k(t)$ has different
convergence times. For doing this, we define a normalised 
probability of $k$ relaxations as
\begin {equation}
p_k(t)={P_k(t)\over P_k(\infty)}~~~.
\end {equation}
Figure 2 shows the plots of $p_k(t)$ vs $t$ for $k=0;1;2;3;4$
for the EW model with $L=10^5$. Note the differences
among the convergences to each steady state: the temporal behaviors of
$p_0$ and $p_1$ quickly go to its
steady values, while $p_2$ and $p_3$ tend to unity only at $t\approx 10^3$ and
$t\approx 10^4$, respectively. This fact suggests that large relaxation
lenghts might occur at large deposition times with small probabilities. So, drawing attention 
to the curve $p_2(t)$ vs $t$, we observe that the saturation occurs
at $t\approx 10^3$. In the RR model with $s=2$, we estimate $t_c\approx 10^3$,
which indicates that the crossover from linear to nonlinear behaviors of the
RR model occurs, for a value of $s$, when $p_{k=s}(t)$ is time independent.

This behavior of $p_k$ is responsible for the dependence of the crossover 
time $t_c$ with the parameter $s$ in the RR model. Figure 3 shows clearly 
this dependence with the plots of $\omega(L,t)/t^{\beta}$ vs $t$
for: (a) $\beta=1/4$ and (b) $\beta=1/3$, with $s=0,2,4$ and $L=10^5$. In (a),
the curve with $s=0$ always grows, while the curve for $s=4$ remains constant,
indicating that $\beta=1/4$, for this value of $L$, is the correct value 
for $s=4$ (EW behavior) 
and a noncorrect for $s=0$ (KK nonlinear behavior). In (b) similar conclusions
are obtained with $\beta=1/3$. For $s=2$, the initial linear and the
asymptotic KPZ behaviors are well observed in (a) and (b), respectively. Note
that the crossover for $s=2$ occurs when the $p_2(t)$ is time independent, that is,
when $p_2(t)\approx 1$, in Figure 2. For $s=4$, we observe only the linear behavior
because the total deposition time is smaller than the crossover time $t_c$.

In order to do a more complete characterization of this crossover, we
also analyse the temporal behavior of the skewness $S(L,t)$.
Figure 4 shows plots of $S(L,t)$ vs $t$ for $L=10^5$ and several values of 
the parameter $s$. For $s=0$, the skewness $S(t)$ goes
quickly to the KPZ transient value $S=-0.28$. As we have explained, $S\neq 0$
means that the interfaces have not up-down symmetry.
We find for $s=2$, a slower approach to the KPZ value, than observed for
$s=0$, indicating that the up-down symmetry is gradually lost when $s>0$.
In particular, for $s=3$, we clearly observe an initial behavior where
$S\approx 0$ and an approach faster to the KPZ value at 
$t\approx 10^4$. The temporal dependence of $S(L,t)$, which is an independent
measurement of the universality class, also indicates an initial 
linear behavior and the KPZ behavior in asymptotic times.
\subsection {$d=2$ Results}
It is also interesting to study this crossover in $d=2$, because changes in the
morphology are expected in this dimension. For the KK model, we have
power-law divergencies of roughness ($\omega^2\sim t^{0.50}$ and 
$\omega^2\sim L^{0.78}$), while the EW model shows logarithmic divergences
($\omega^2\sim\ln t$ and $\omega^2\sim\ln L$.

In order to avoid saturation effects, we work with $L=2000$ (4$\times 10^6$ sites),
and we do simulations until $t=10^4$. Due to the computational cost,
we perform only two samples for each value of the parameter $s$ and,
consequently, the data quality in this section is poorer than in the previous
subsection. The crossover is analysed through a similar process done
in $d=1$.
We expect logarithmic behavior when the linear term dominates, so we define
\begin{equation}
Y_s(t)={\omega^2(t)-B_s\over A_s\ln(t)}~~,
\end{equation}
where $A_s$ and $B_s$ are the
$s$-dependent coefficients obtained through a logarithmic regression in the
interval $2<t<100$.
In Figure 5, we plot the temporal behavior of the function $Y_s(t)$, 
in a semi-logarithmic scale.
If the temporal behavior of the roughness is 
logarithmic, $Y_s(t)$ must be a constant, equal to one.
We note that this occurs for $s=\infty$, but, for $s=1$, this
constant behavior remains until $t\approx 10^2$. After this time, there is
a crossover to the power law behavior. However, until the studied time, it is not
possible to determine the exponent.

In order to verify the existence of a power law behavior with the KPZ
exponent $\beta$, we need to analyse values 
of the parameter $s$ smaller than $s=1$, that is,
we need to do continuum variations in $s$. So, we do this by assinging a
probability $s$ for the particle relax one lattice unit and a probability
$(1-s)$ for the particle be evaporated.
In figure 6, we show the graph of $\omega/t^\beta$ vs $t$, for $s=0$ (KK model) and 
$s=0.1,0.3,0.5$. In this graph, it was used $\beta=1/4$ which is the  
exponent for the KK model in $d=2$ \cite {kk}.
If the temporal behavior of the roughness has a power law behavior with this 
exponent, the curves must be horizontal.
For the RR model ($s\neq 0$), the curves show asymptotic approaches to this nonlinear 
behavior. For greater values of the parameter $s$, the crossover time to this
behavior occurs for times greater than the time studied ($t=10^4$). 
\section {Conclusions}
We have studied the model with restricted surface
relaxation which combines the main features of the model with
surface relaxation (EW model) and the model with refuse (KK model). 
A power-law temporal behavior of roughness with two growth exponents 
was observed in $d=1$: The linear growth exponent, $\beta=1/4$, occurs in short
times and the nonlinear one, $\beta=1/3$, appears in the asymptotic limit. 
This result suggests the following description: the linear
term of the KPZ equation (Eq.7) dominates in short times and the nonlinear 
term dominates in asymptotic times. We also noted that the crossover time 
$t_c$ is independent of the system
size $L$ and it is only function of the parameter $s$.
This result corroborates the
renormalization group solution made by Nattermann and Tang \cite {nattang},
where the KPZ equation with small nonlinear term was considered.
In $d=2$, we have found indications of the same kind of crossover:
A logarithmic temporal behavior of roughness in short times,
which is related to the linear EW equation, and 
a power-law behavior with $\beta= 0.25$ in asymptotic times, related to 
the nonlinear KPZ equation.
\bigskip

\centerline {\bf Acknowledgments}
The authors would like to thank Rog\'erio Magalh\~aes Paniago for a critical 
reading of the manuscript. 
The simulations were made at an ensemble of Digital Alpha 500 Au of the
Departamento de F\'{\i}sica -UFMG and at a Sun HPC 10000 of the CENAPAD 
MG-CO. 
This work was supported by CNPq, Fapemig and Finep/Pronex, Brazilian agencies.  
                                                                              
%
%
%
%
%

\newpage

{\large \bf Figure captions}

\bigskip

{\bf Figure 1}

The temporal behavior of the roughness $\omega(L,t)$ for the RR model 
simulated at a substrate with $L=10^5$ sites and $s=2$, in a log-log plot. 
The two straight lines are showing the
power-law fit results with $\beta=1/4$ (short times) and $\beta=1/3$
(asymptotic times). The crossover time $t_c$ is defined as the intersection 
of these two lines, as indicated in the figure.

{\bf Figure 2}

The temporal behavior of normalised probability of $k$ relaxations
$p_k(t)$ for the EW model with $L=10^5$ sites for $k=0;1;2;3;4$, from top to below.

{\bf Figure 3}

The log-log plots of temporal behaviors of: (a) $\omega/t^{1/4}$
and (b)$\omega/t^{1/3}$ for $s=0;2;4$ and $L=10^5$. 
We might observe, for $s=2$, when the
system drives away from the linear behavior in (a) and,  
the arrival at the nonlinear KPZ behavior in (b). For $s=4$, the
system must approach to the KPZ behavior at deposition time
greater than $t=10^5$. 

{\bf Figure 4}

The temporal behavior of the skewness $S(L,t)$ for $L=10^5$ for
several values of the parameter $s$. The two horizontal
lines show the value $S=0$ (EW value) and
$S=-0.28$ (KPZ value).

{\bf Figure 5}

The temporal behavior, in a semi-logarithmic scale, of the function
$Y_s(t)$ for $s=1$ and $s=\infty$ (EW) with $L=2000$. The dashed
horizontal line indicates the constant value $Y_s(t)=1$.

{\bf Figure 6}

The semi-logarithmic graph of $\omega(L,t)/t^{1/4}$ in function of
the time $t$ for $L=2000$ and some values of the parameter $s$.

\newpage
\centerline {TABLE 1}
\begin{center}
\begin{tabular}{lllllll}  \hline \hline
$k$ &~~~~~~~~$P_k(\infty)$  \\ \hline
$0$    &~~~~~~~0.5813(3)\\
$1$    &~~~~~~~0.3481(3)\\
$2$    &~~~~~~~0.0616(1)\\
$3$    &~~~~~~~0.00812(2)\\
$4$    &~~~~~~~0.00085(1)\\
$5$    &~~~~~~~7(5)$\times 10^{-5}$\\
\hline \hline
\end {tabular}
\end {center}
The steady state values of probability of diffusion $k$ sites determined as
the mean value in the interval $10^4\leq t\leq 10^5$ for a EW system with
$L=10^5$.


\begin {thebibliography}{99}
%
\bibitem {meakin}
P. Meakin, 
\newblock {\it Fractals, Scaling and Growth Far from Equilibrium},
\newblock Cambridge Univ. Press, Cambridge (1998).

\bibitem {barab}
A.-L. Barab\'asi and H. E. Stanley, 
\newblock {\it Fractal Concepts in Surface Growth},
\newblock Cambridge Univ. Press, Cambridge (1995)

\bibitem {krug1}
J. Krug,
\newblock Adv. Phys. {\bf 46}, 139 (1997).

\bibitem {db}
M. J. Vold, 
\newblock J. Phys. Chem. {\bf 64}, 1616 (1960).
P. Meakin, P. Ramanlal, L. M. Sander and R. C. Ball,
\newblock Phys. Rev. A {\bf 34}, 5091 (1986).

\bibitem {eden}
M. Eden,
\newblock Proc. Fourth Berkeley Symp. Mathematical Statistics and
Probability, Volume IV. Edited by L. Le Cam and J. Neyman: 
Biology and Problems of Health (University of
California Press, Berkeley, 1961).

\bibitem {ew}
S. F. Edwards and D. R. Wilkinson,
\newblock Proc. R. Soc. A {\bf 381}, 17 (1982).

\bibitem {family86}
F. Family,
\newblock J. Phys A {\bf 19}, L441 (1986).

\bibitem {kk}
J. M. Kim and J. M. Kosterlitz,
\newblock Phys. Rev. Lett. {\bf 62}, 2289 (1989).

\bibitem {wolfvillain}
D. E. Wolf and J. Villain,
\newblock Europhys. Lett {\bf 13}, 389 (1990).

\bibitem {fv}
F. Family and T. Vicsek,
\newblock J. Phys. A {\bf 18}, L75 (1985).

\bibitem {kpz}
M. Kardar, G. Parisi and Y.-C. Zhang,
\newblock Phys. Rev. Lett. {\bf 56}, 889 (1886).

\bibitem {moser}
K. Moser, J. Kert\'esz and D. E. Wolf,
\newblock Physica A {\bf 178}, 215 (1991).

\bibitem {ggg}
B. Grossmann, H. Guo and M. Grant,
\newblock Phys. Rev. A {\bf 43}, 1727 (1991).

\bibitem {yks}
H. Yan, D. Kessler and L. M. Sander,
\newblock Phys. Rev. Lett. {\bf 64}, 926 (1990).

\bibitem {pelle}
Y. P. Pellegrini and R. Jullien,
\newblock Phys. Rev. Lett {\bf 64}, 1745 (1990);
\newblock Phys. Rev. A {\bf 43} 920 (1991).

\bibitem {nattang}
T. Nattermann and L.-H. Tang,
\newblock Phys. Rev. A {\bf 45}, 7156 (1992).

\bibitem {krugskew}
J. Krug, P. Meakin and T. Halpin-Healy,
\newblock Phys. Rev. A {\bf 45}, 638 (1992).

\bibitem {landau2}
Y. Shim and D. P. Landau,
\newblock Phys. Rev. E {\bf 64} No 036110 (2001).

\bibitem {skew}
C.-S. Chin and M. den Nijs,
\newblock Phys. Rev. E {\bf 59}, 2633 (1999);
M. Pr\"ahofer and H. Spohn,
\newblock Phys. Rev. Lett. {\bf 84}, 4882 (2000);

\bibitem {krugmeakin}
J. Krug and P. Meakin,
\newblock J. Phys. A {\bf 23}, L987 (1990).

\bibitem{ew2d}
S. Pal and D. P. Landau,
\newblock Physica A {\bf 267}, 406 (1999).

\end {thebibliography}
\end {document}